\documentclass[twocolumn]{aastex631}

\newcommand{\Ks}{K_{\rm s}}
\newcommand{\kms}{{{\rm km~s}^{-1}}}
\newcommand{\Gaia}{{{\it Gaia}}}
\newcommand{\WISE}{{{\it WISE}}}
\newcommand{\NEOWISE}{{{\it NEOWISE}}}
\newcommand{\Spitzer}{{{\it Spitzer}}}
\newcommand{\Teff}{T_{\rm eff}}
\newcommand{\FeH}{{\rm [Fe/H]}}
\newcommand{\Ic}{{I_\mathrm{c}}}
\newcommand{\rmin}{{r_\mathrm{min}}}
\newcommand{\rmax}{{r_\mathrm{max}}}
\newcommand{\zmax}{{z_\mathrm{max}}}

\received{October 11, 2021}
\revised{October 27, 2021}
\accepted{October 27, 2021}
\submitjournal{ApJ}

\shorttitle{A very metal-poor RR Lyrae with a disk orbit}
\shortauthors{Matsunaga et al.}
\graphicspath{{./}{figures/}}
\begin{document}

\title{A Very Metal-poor RR Lyrae Star with a Disk Orbit Found in the Solar Neighborhood}

\correspondingauthor{Noriyuki Matsunaga}
\email{matsunaga@astron.s.u-tokyo.ac.jp}

\author{Noriyuki Matsunaga}
\affiliation{Department of Astronomy, School of Science, The University of Tokyo, 7-3-1 Hongo, Bunkyo-ku, Tokyo 113-0033, Japan}
\affiliation{Laboratory of Infrared High-resolution spectroscopy (LiH), Koyama Astronomical Observatory, Kyoto Sangyo University, Motoyama, Kamigamo, Kita-ku, Kyoto, 603-8555, Japan}
\author{Akinori Itane}
\affiliation{Department of Astronomy, School of Science, The University of Tokyo, 7-3-1 Hongo, Bunkyo-ku, Tokyo 113-0033, Japan}
\author[0000-0001-6924-8862]{Kohei Hattori}
\affiliation{National Astronomical Observatory of Japan, 2-21-1 Osawa, Mitaka, Tokyo 181-8588, Japan}
\affiliation{Institute of Statistical Mathematics, 10-3 Midoricho, Tachikawa, Tokyo 190-0014, Japan}
\author[0000-0001-8926-3496]{Juliana Crestani}
\affiliation{Dipartimento di Fisica, Universita di Roma Tor Vergata, via della Ricerca Scientifica 1, I-00133 Roma, Italy}
\affiliation{INAF, Osservatorio Astronomico di Roma, via Frascati 33, I-00078 Monte Porzio Catone, Italy}
\author[0000-0001-7511-2830]{Vittorio Braga}
\affiliation{INAF, Osservatorio Astronomico di Roma, via Frascati 33, I-00078 Monte Porzio Catone, Italy}
\affiliation{Space Science Data Center, ASI, via del Politecnico snc, I-00133 Roma, Italy}
\author[0000-0002-4896-8841]{Giuseppe Bono}
\affiliation{Dipartimento di Fisica, Universita di Roma Tor Vergata, via della Ricerca Scientifica 1, I-00133 Roma, Italy}
\affiliation{INAF, Osservatorio Astronomico di Roma, via Frascati 33, I-00078 Monte Porzio Catone, Italy}
\author[0000-0002-2861-4069]{Daisuke Taniguchi}
\affiliation{Department of Astronomy, School of Science, The University of Tokyo, 7-3-1 Hongo, Bunkyo-ku, Tokyo 113-0033, Japan}
\author[0000-0002-2154-8740]{Junichi Baba}
\affiliation{National Astronomical Observatory of Japan, 2-21-1 Osawa, Mitaka, Tokyo 181-8588, Japan}
\author[0000-0003-0332-0811]{Hiroyuki Maehara}
\affiliation{National Astronomical Observatory of Japan, 2-21-1 Osawa, Mitaka, Tokyo 181-8588, Japan}
\author{Nobuharu Ukita}
\affiliation{National Astronomical Observatory of Japan, 2-21-1 Osawa, Mitaka, Tokyo 181-8588, Japan}
\author{Tsuyoshi Sakamoto}
\affiliation{Smile Collection Inc., 4-13-10 Kotobuki, Taito-ku, Tokyo 111-0042, Japan}
\author[0000-0003-4578-2619]{Naoto Kobayashi}
\affiliation{Institute of Astronomy, School of Science, The University of Tokyo, 2-21-1 Osawa, Mitaka, Tokyo 181-0015, Japan}
\affiliation{Kiso Observatory, Institute of Astronomy, School of Science, The University of Tokyo, 10762-30 Mitake, Kiso-machi, Kiso-gun, Nagano 397-0101, Japan}
\affiliation{Laboratory of Infrared High-resolution spectroscopy (LiH), Koyama Astronomical Observatory, Kyoto Sangyo University, Motoyama, Kamigamo, Kita-ku, Kyoto, 603-8555, Japan}
\author{Tsutomu Aoki}
\affiliation{Kiso Observatory, Institute of Astronomy, School of Science, The University of Tokyo, 10762-30 Mitake, Kiso-machi, Kiso-gun, Nagano 397-0101, Japan}
\author{Takao Soyano}
\affiliation{Kiso Observatory, Institute of Astronomy, School of Science, The University of Tokyo, 10762-30 Mitake, Kiso-machi, Kiso-gun, Nagano 397-0101, Japan}
\author{Ken'ichi Tarusawa}
\affiliation{Kiso Observatory, Institute of Astronomy, School of Science, The University of Tokyo, 10762-30 Mitake, Kiso-machi, Kiso-gun, Nagano 397-0101, Japan}
\author{Yuki Sarugaku}
\affiliation{Laboratory of Infrared High-resolution spectroscopy (LiH), Koyama Astronomical Observatory, Kyoto Sangyo University, Motoyama, Kamigamo, Kita-ku, Kyoto, 603-8555, Japan}
\author{Hiroyuki Mito}
\affiliation{UTokyo Organization for Planetary and Space Science, the University of Tokyo, 7-3-1, Hongo, Tokyo 113-0033, Japan}
\author[0000-0002-8792-2205]{Shigeyuki Sako}
\affiliation{Institute of Astronomy, School of Science, The University of Tokyo, 2-21-1 Osawa, Mitaka, Tokyo 181-0015, Japan}
\author{Mamoru Doi}
\affiliation{Institute of Astronomy, School of Science, The University of Tokyo, 2-21-1 Osawa, Mitaka, Tokyo 181-0015, Japan}
\author{Yoshikazu Nakada}
\affiliation{Institute of Astronomy, School of Science, The University of Tokyo, 2-21-1 Osawa, Mitaka, Tokyo 181-0015, Japan}
\author[0000-0003-1604-9127]{Natsuko Izumi}
\affiliation{Institute of Astronomy and Astrophysics, Academia Sinica, No.~1, Section~4, Roosevelt Road, Taipei 10617, Taiwan}
\author{Yoshifusa Ita}
\affiliation{Astronomical Institute, Graduate School of Science, Tohoku University, 6-3 Aramaki Aoba, Aoba-ku, Sendai, Miyagi 980-8578, Japan}
\author[0000-0002-8015-0476]{Hiroki Onozato}
\affiliation{National Astronomical Observatory of Japan, 2-21-1 Osawa, Mitaka, Tokyo 181-8588, Japan}
\author[0000-0002-5649-7461]{Mingjie Jian}
\affiliation{Department of Astronomy, School of Science, The University of Tokyo, 7-3-1 Hongo, Bunkyo-ku, Tokyo 113-0033, Japan}
\author{Sohei Kondo}
\affiliation{Kiso Observatory, Institute of Astronomy, School of Science, The University of Tokyo, 10762-30 Mitake, Kiso-machi, Kiso-gun, Nagano 397-0101, Japan}
\author[0000-0002-6505-3395]{Satoshi Hamano}
\affiliation{National Astronomical Observatory of Japan, 2-21-1 Osawa, Mitaka, Tokyo 181-8588, Japan}
\author[0000-0003-3579-7454]{Chikako Yasui}
\affiliation{National Astronomical Observatory of Japan, California Office, 100 W. Walnut St., Suite 300, Pasadena, CA 91124, USA}
\author[0000-0002-9397-3658]{Takuji Tsujimoto}
\affiliation{National Astronomical Observatory of Japan, 2-21-1 Osawa, Mitaka, Tokyo 181-8588, Japan}
\author{Shogo Otsubo}
\affiliation{Laboratory of Infrared High-resolution spectroscopy (LiH), Koyama Astronomical Observatory, Kyoto Sangyo University, Motoyama, Kamigamo, Kita-ku, Kyoto, 603-8555, Japan}
\author[0000-0003-2380-8582]{Yuji Ikeda}
\affiliation{Photocoding, 460-102 Iwakura-Nakamachi, Sakyo-ku, Kyoto 606-0025, Japan}
\affiliation{Laboratory of Infrared High-resolution spectroscopy (LiH), Koyama Astronomical Observatory, Kyoto Sangyo University, Motoyama, Kamigamo, Kita-ku, Kyoto, 603-8555, Japan}
\author{Hideyo Kawakita}
\affiliation{Laboratory of Infrared High-resolution spectroscopy (LiH), Koyama Astronomical Observatory, Kyoto Sangyo University, Motoyama, Kamigamo, Kita-ku, Kyoto, 603-8555, Japan}
\affiliation{Department of Astrophysics and Atmospheric Sciences, Faculty of Science, Kyoto Sangyo University, Motoyama, Kamigamo, Kita-ku, Kyoto 603-8555, Japan}

\begin{abstract}
Metal-deficient stars are important tracers for
understanding the early formation of the Galaxy.
Recent large-scale surveys with both photometric and spectroscopic data
have reported an increasing number of metal-deficient stars whose kinematic features
are consistent with those of the disk stellar populations. 
We report the discovery of an RR~Lyrae variable (hereafter RRL)
that is located within
the thick disk
and has an orbit consistent with the thick-disk kinematics.
Our target RRL (HD\,331986) is located at around 1\,kpc from the Sun and, 
with $V \simeq 11.3$
, is among the {$\sim$}130 brightest RRLs known so far.
However, this object was scarcely studied because it is in the midplane
of the Galaxy, the Galactic latitude around $-1 \deg$.
Its near-infrared spectrum (0.91--1.32\,{$\mu$}m) shows no absorption line
except hydrogen lines of the Paschen series, suggesting
$\FeH \lesssim -2.5$.
It is the most metal-deficient RRL, at least, among 
the RRLs whose orbits are consistent with the disk kinematics,
although we cannot determine to which of the disk and the halo
it belongs.
This unique RRL would provide us with essential clues for
studying the early formation of stars in the inner Galaxy
with further investigations, including high-resolution optical spectroscopy.
\end{abstract}

\keywords{RR Lyrae variable stars (1410), Spectroscopy (1558), Milky Way disk (1050), Surveys (1671), Milky Way formation (1053), Metallicity (1031)}

\section{Introduction} \label{sec:intro}
It is possible to investigate stellar populations in the Galaxy
with various detailed observations thanks to relatively low
distances to individual stars.
For example, the positions and motions of billions of stars
provided by the {\Gaia} satellite 
\citep[see][for the latest data release, EDR3]{Gaia-2021}
have revolutionized
our understanding of the Galactic structure and evolution. 
In particular, streams and substructures that are clearly 
detected with the {\Gaia} data allow us to build up a scenario
of accretions and mergers in the early history of the Galaxy
\citep{Helmi-2020}.
Another vital material for characterizing
the Galactic stellar populations is being brought by 
detailed elemental abundances based on high-resolution spectroscopy
\citep[][and references therein]{Jofre-2019,Matteucci-2021}.
Metal-deficient stars are especially important for
studying the early Galactic evolution \citep{Frebel-2015},
and tremendous efforts have been devoted
to identifying metal-deficient stars
\citep[see, e.g.,][for recent large-scale surveys]{Starkenburg-2017,DaCosta-2019}. 
An exciting discovery through these surveys is
the presence of metal-deficient stars in the Galactic disk \citep{Sestito-2020}.

The presence and the characteristics of metal-deficient stars in the disk
provide us with crucial clues to the formation of the Galactic disk
and its history. A strong merger at an early stage, for example,
would have disrupted
the disk, and metal-deficient stars that were 
present at the time of the merger could lose their disk-like kinematics
except those in 
the inner Galaxy where the gravitational potential was deep enough
to trap the stars. 
Recent studies equipped with the {\Gaia} data suggest
that such a merger occurred {$\sim$}10\,Gyr ago,
triggered the growth of
the thick disk \citep{Gallart-2019,Helmi-2018,Helmi-2020},
and is imprinted in the kinematics of halo stars as a feature
called {\it Gaia Enceladus} or {\it Gaia Sausage}.
The metal-deficient stars with disk-like orbits
such as the stars with $\FeH \leq -2.5$
discussed in \citet{Sestito-2020}
can be considered relics of the proto-disk that
was present before the merger.
Further identification of such stars over a wide range of metallicity,
together with the characterization of halo stars in the same spatial volume, 
is crucial to establish (or reject) such a formation scenario
of the thick disk. 

This study focuses on RR~Lyrae stars  (RRLs)
to investigate the old and metal-deficient populations
located in the Galactic disk. 
RRLs are pulsating stars in the Cepheid instability strip
and at the horizontal branch phase
evolved from low-mass stars ($\lesssim 1\,M_\odot$). 
They are exclusively old ($> 10$\,Gyr) and thus trace
old stellar populations in galaxies. 
General views on the characteristics of RRLs are found in
a review by \citet{Beaton-2018} and references therein.
The majority of RRLs belong to old stellar spheroids,
i.e., the halo and the bulge, 
while the disk population of RRLs has been also found
\citep{Layden-1995,Prudil-2020,Zinn-2020}. 
Other than the differences in spatial distribution and kinematics,
an essential difference between the halo and disk groups of RRLs
is the metallicity distribution. Disk RRLs are predominantly
metal-rich, $\FeH \gtrsim -1$, while halo RRLs are less metal enriched.
In this study, we report the discovery of an RRL with
$\FeH \lesssim -2.5$ that has a disk-like orbit.  

The rest of this paper is organized as follows.
First, we discuss the photometric properties of
our target in Section~\ref{sec:photometric}.
We first identified this object as 
a bright but unexplored RRL through our variability survey,
KWFC Intensive Survey of the Galactic Plane (KISOGP), described 
in Section~\ref{subsec:KISOGP}. 
Combined with other photometric data (Section~\ref{subsec:other-photo})
and the {\it Gaia}-based distance, we give an estimate of its metallicity
by making use of the period--luminosity--metallicity relation
of RRLs (Section~\ref{subsec:PLR}). 
Then, in Section~\ref{sec:spectroscopic}, we present the analysis
of the near-infrared spectrum obtained with the WINERED spectrograph. 
We detected hydrogen lines of the Paschen series, which enable us
to measure the radial velocity (Section~\ref{subsec:Hlines}), 
but detected no metallicity lines, which gives only upper limits
of abundance (Section~\ref{subsec:metallic}). 
In Section~\ref{sec:discussion}, we discuss
the kinematics and the low metallicity of the target RRL
in the context of the Galactic structure and evolution. 
Finally, Section~\ref{sec:conclusion} concludes the paper.

\section{Photometric data} \label{sec:photometric}
\subsection{KISOGP} \label{subsec:KISOGP}
KISOGP is the large-scale survey of variable stars
in the northern Galactic plane using
Kiso Wide Field Camera (KWFC) attached to the 105-cm
Schmidt telescope at Kiso Observatory, Japan.
KWFC is a mosaic CCD camera with eight CCD chips
having a total of $8\,{\rm k} \times 8\,{\rm k}$ pixels
covering a field-of-view of 2.2~degrees square
(0.946$^{\prime\prime}$\,pix$^{-1}$) on the sky.
See more details of the KWFC in \citet{Sako-2012}. 

To discover and characterize variable stars
in the Galactic plane, we started the KISOGP
in 2012 and made $\Ic$-band time-series observations
for the 80 KWFC fields-of-view covering
${\sim}330\,\deg ^2$ between 60 and 210\,$\deg$ in Galactic longitude \citep{Matsunaga-2017}. 
The analysis for publishing the catalog of
variables detected in the KISOGP is in progress,
but a study on eclipsing binary systems has been 
published by \citet{Ren-2021}.

\begin{figure}
\begin{center}
\includegraphics[width=0.6\hsize]{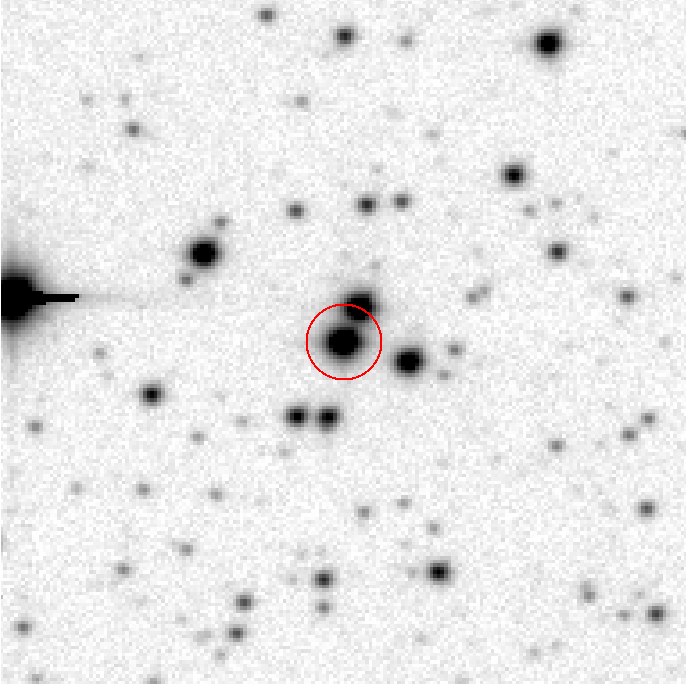}
\end{center}
\caption{
Finding chart of our target RRL (HD\,331986), indicated by the red circle with
the radius of 10\,\arcsec, on a KISOGP $I$-band image
(3 arc-minute square, north up and east left). 
\label{fig:chart}}
\end{figure}

\begin{figure}
\begin{center}
\includegraphics[width=0.8\hsize]{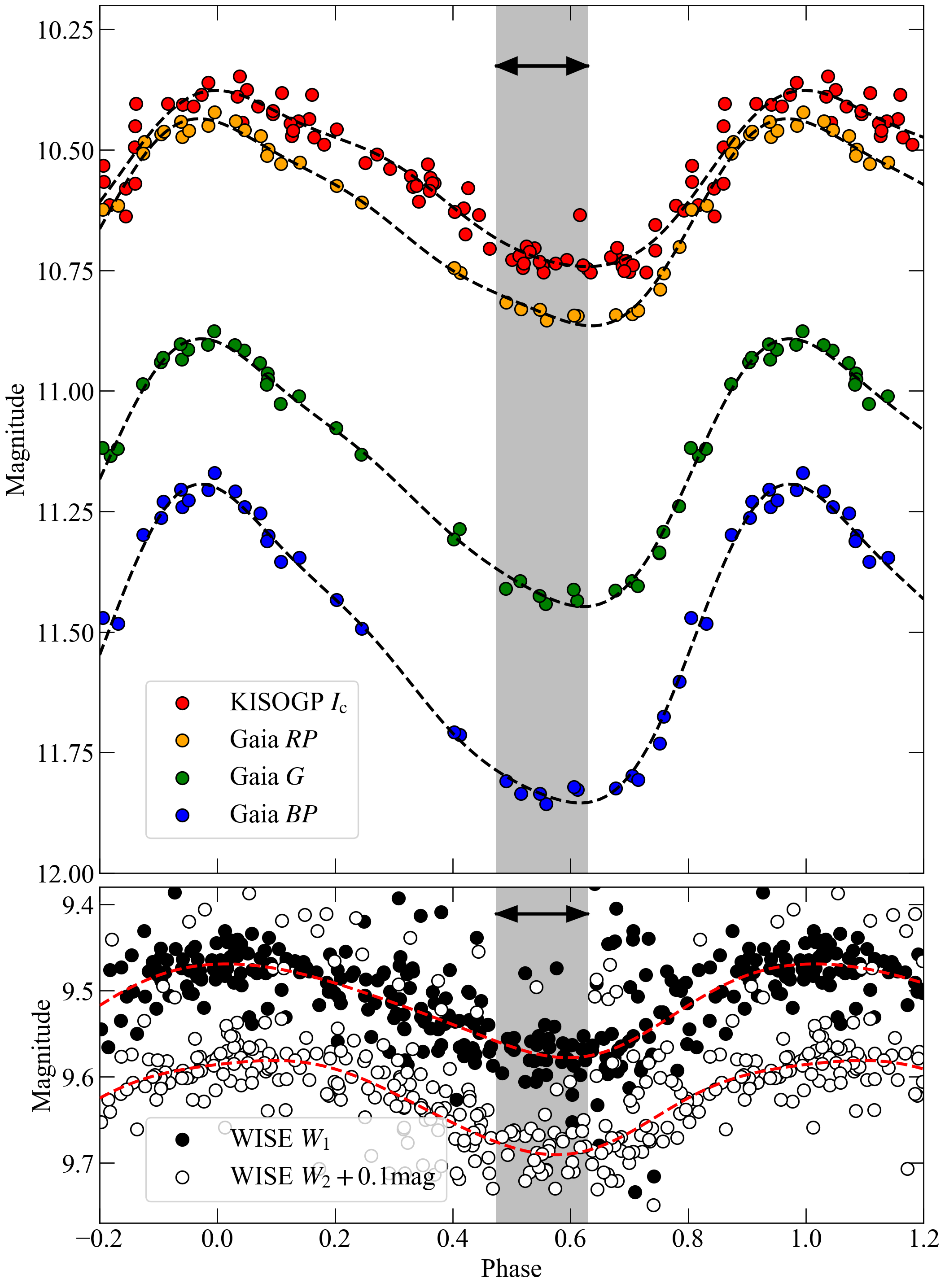}
\end{center}
\caption{
Light curves of our target RRL (HD\,331986). The upper panel presents optical data,
i.e., KISOGP ($\Ic$) and {\Gaia} DR2 ($RP$, $G$, and $BP$),
while the lower panel presents the infrared data
from {\NEOWISE} ($W_1$ and $W_2$). The dashed curves indicate
the fitted discrete Fourier series (Equation~\ref{eq:fourier}).
The gray strip indicates the duration of
the WINERED spectroscopic observation (Section~\ref{subsec:WINERED}).
\label{fig:LCs}}
\end{figure}

During our early attempt to identify periodic variables,
we discovered a bright but scarcely-investigated
RRL and made a spectroscopic observation in 2015 (Section~\ref{subsec:WINERED}).
The properties of this RRL variable,
HD\,331986 (finding chart presented in Figure~\ref{fig:chart}),
are summarized in Table~\ref{tab:obj}.
The KISOGP $I$-band light curve of this object
is presented in Figure~\ref{fig:LCs}
together with light curves in other wavelengths
(see Section~\ref{subsec:other-photo}).
\begin{deluxetable}{lll}
%% \tablenum{1}
\tablecaption{Properties of the target RR~Lyr star\label{tab:obj}}
\tablewidth{0pt}
\tablehead{
\colhead{Item(s)} & \colhead{Value(s)} & \colhead{Ref.} \\
}
\startdata
KISOGP ID & KISOJ\,201241.60$+$321242.4 & 1 \\
Aliases & HD\,331986, NSVS\,8487853 & 2 \\
{\Gaia} EDR3 Source ID & 2054159819759156992 & 3 \\
Eq.\  coordinate ($\deg$) & $\alpha=303.17334, ~\delta=32.21177$ & 2 \\
Gal.\  coordinate ($\deg$) & $l=70.40915, ~b=-1.05159$ & 2 \\
Valiability type & RRc & 1, 2 \\
Period (days) & 0.371197 & 1 \\
2MASS magnitudes & $J=9.954$, ~$H=9.663$ & 4 \\
 & $\Ks=9.577$ &  \\
Astrometric distance (kpc) & $1.042 \pm 0.015$ & 5 \\
Proper mosion (mas~yr$^{-1}$) & $\mu_\alpha \cos\delta = 11.76, ~\mu_\delta = -5.09$ & 3 \\
 & $\mu_l \cos b = 2.23, ~\mu_b = -12.62$ & 3 \\
\enddata
\tablerefs{(1) This work, (2) Simbad, (3) {\Gaia} EDR3, (4) \citet{Skrutskie-2006}, (5) \citet{BailerJones-2021}}
\end{deluxetable}

We fitted the discrete Fourier series to the photometric points, 
\begin{equation}
m(t) = A_0 + \sum_{j=1}^{3} A_j \sin \left[ \frac{2\pi j}{P} (t-t_0) + \phi_j\right]
\label{eq:fourier}
\end{equation}
where $t$ indicates the heliocentric Julian date (HJD)
of each photometric measurement
and $t_0$ is the reference epoch, 2456000.519, which we determined 
to put the maximum of the $\Ic$-band light curve at phase zero. 
We obtained the period, $P=0.371197$\,days, that gives the maximum power
in the periodogram constructed with the three-term Fourier model
as given in Equation~\ref{eq:fourier} for the KISOGP
light curve \citep[see the methodology, e.g., in][]{VanderPlas-2018}.
Using the fitted light curve, we calculated 
the intensity mean and the peak-to-valley amplitude
(Table~\ref{tab:means}).
We confirm that this star is a first-overtone mode RRL (RRc type)
based on its period, amplitude, and light curve shape.
\begin{deluxetable}{ccccccc}
%% \tablenum{2}
\tablecaption{Mean magnitudes and peak-to-valley amplitudes\label{tab:means}}
\tablewidth{0pt}
\tablehead{
Band & \colhead{$\Ic$} & \colhead{$RP$} & \colhead{$G$} & \colhead{$BP$} & \colhead{$W_1$} & \colhead{$W_2$}
}
\startdata
Mean & 10.556 & 10.648 & 11.165 & 11.525 & 9.519 & 9.529 \\
Amp  & 0.365 & 0.430 & 0.556 & 0.661 & 0.109 & 0.109 \\
\enddata
\end{deluxetable}

\subsection{Other photometric data sets} \label{subsec:other-photo}
Our target star was first identified as a candidate of RRL by 
\citet{Kinemuchi-2006} based on the Northern Sky Variability Survey (NSVS).
\citet{Hoffman-2009} also reported this star as an RRL
based on the automated classification with the NSVS data. 
It has also been included as an RRL 
in the {\Gaia} Data Release 2 \citep{Clementini-2019} and 
the variability catalog based on
the {\it Wide-field Infrared Survey Explorer}
({\WISE}) \citep{Chen-2018}.
However, no detailed follow-up study has been done,
and its kinematic and chemical features remained to be revealed.

As seen in Figure~\ref{fig:chart},
there is a similarly bright star at {$\sim$}10\arcsec,
{\Gaia} EDR3 ID~2054159819759157504 with $G=11.56$.
This star is at 0.35\,kpc \citep{BailerJones-2021}
and has a higher proper motion {$\sim$}20\,mas\,yr{$^{-1}$}.
It does not affect the KISOGP photometry of the target RRL, but
severe contamination occurs in the NSVS images
with the pixel size of 14.{\arcsec}4 \citep{Kinemuchi-2006}. %% \citep{Wozniak-2004}, 
The same applies to the data from
the All-Sky Automated Survey for Supernovae
(ASAS-SN) with 8.{\arcsec}0 pixels \citep{Jayasinghe-2018}. 
There is no optical data set, other than the {\Gaia} DR2, 
that was published before and gives a high-quality light curve
of this object in $V$ or other bands.

We consider the data of 2MASS \citep{Skrutskie-2006} and
{\NEOWISE} \citep{Mainzer-2011,Mainzer-2014} for the infrared range. 
While the 2MASS gives the $JH\Ks$ magnitudes at a single epoch
(1999 June 22), {\NEOWISE} gives time-series data collected
between 2014 May and 2020 October in the two mid-IR bands,
$W_1$ (3.4\,{$\mu$}m) and $W_2$ (4.6\,{$\mu$}m).
The available time-series data are presented in 
Figure~\ref{fig:LCs}, but the 2MASS magnitudes are not 
included because we cannot determine the precise pulsation phase
for this data set $\sim$30 years ago.
We fitted the three-term Fourier series (Equation~\ref{eq:fourier}) to
the photometric points in each band and estimated
the intensity mean and amplitude (Table~\ref{tab:means}).
The {\NEOWISE} data include many outliers,
and we made 2-{$\sigma$} clipping for fitting the Fourier series. 
Although we did not find good $V$-band photometry 
free from the blending effect as mentioned above,
the mean magnitudes in
the {\Gaia} bands give the mean $V$-band mean magnitude of 11.32
and the maximum magnitude of 11.01
according to the formula in \citet{Evans-2018}.
\citet{Maintz-2005} compiled a catalog of 
well-identified RRLs brighter than $V=12.5$
at the maximum phase, and there are 132 RRLs 
brighter than our target RRL. 

\subsection{Period--luminosity--metallicity relation} \label{subsec:PLR}
RRLs are established distance indicators although
their correlation between period and absolute magnitudes 
are significantly affected by metallicity \citep{Beaton-2018}.
We consider the period--luminosity--metallicity (PLZ) relations
obtained by \citet{Neeley-2019} to see if the photometric data
are consistent with the geometric distance and to give
a constraint on the metallicity of our target.
Following a theoretical study on the PLZ relation in \citet{Neeley-2017},
\citet{Neeley-2019} used the {\Gaia} DR2 trigonometric distances of 55 RRLs
with [Fe/H] between $-2.56$ and $-0.07$
to obtain the empirical PLZ relations from optical to
mid-IR photometric bands. The PLZ relation in each band is in the form of
\begin{equation}
M_\lambda = a + b (\log P_\mathrm{F}+0.30) + c (\FeH+1.36),
\label{eq:PLZ}
\end{equation}
where $P_\mathrm{F}$ is the ``fundamentalized'' period
given by $\log P_\mathrm{F} = \log P + 0.127$
for an RRc star ($\log P_\mathrm{F}=-0.303$ for our target).
The mid-IR bands used in \citet{Neeley-2019} are those of
the {\Spitzer} Space Telescope, but we use their PLR relations for
the {\WISE} data. 
The {\Spitzer} [3.6] and [4.5] bands correspond to
the {\WISE} $W_1$ and $W_2$ bands, respectively. 
The theoretical result by \citet{Neeley-2017} suggests that
the relations in the {\Spitzer} bands and those in the {\WISE} bands
identical, within 0.003\,mag, with each other at each wavelength.
In addition to $W_1$, $W_2$, and $I_\mathrm{c}$ in Table~\ref{tab:means}, 
we consider the single-epoch 2MASS magnitudes in $JH\Ks$
in the following analysis.

\begin{figure}
\begin{center}
\includegraphics[width=0.9\hsize]{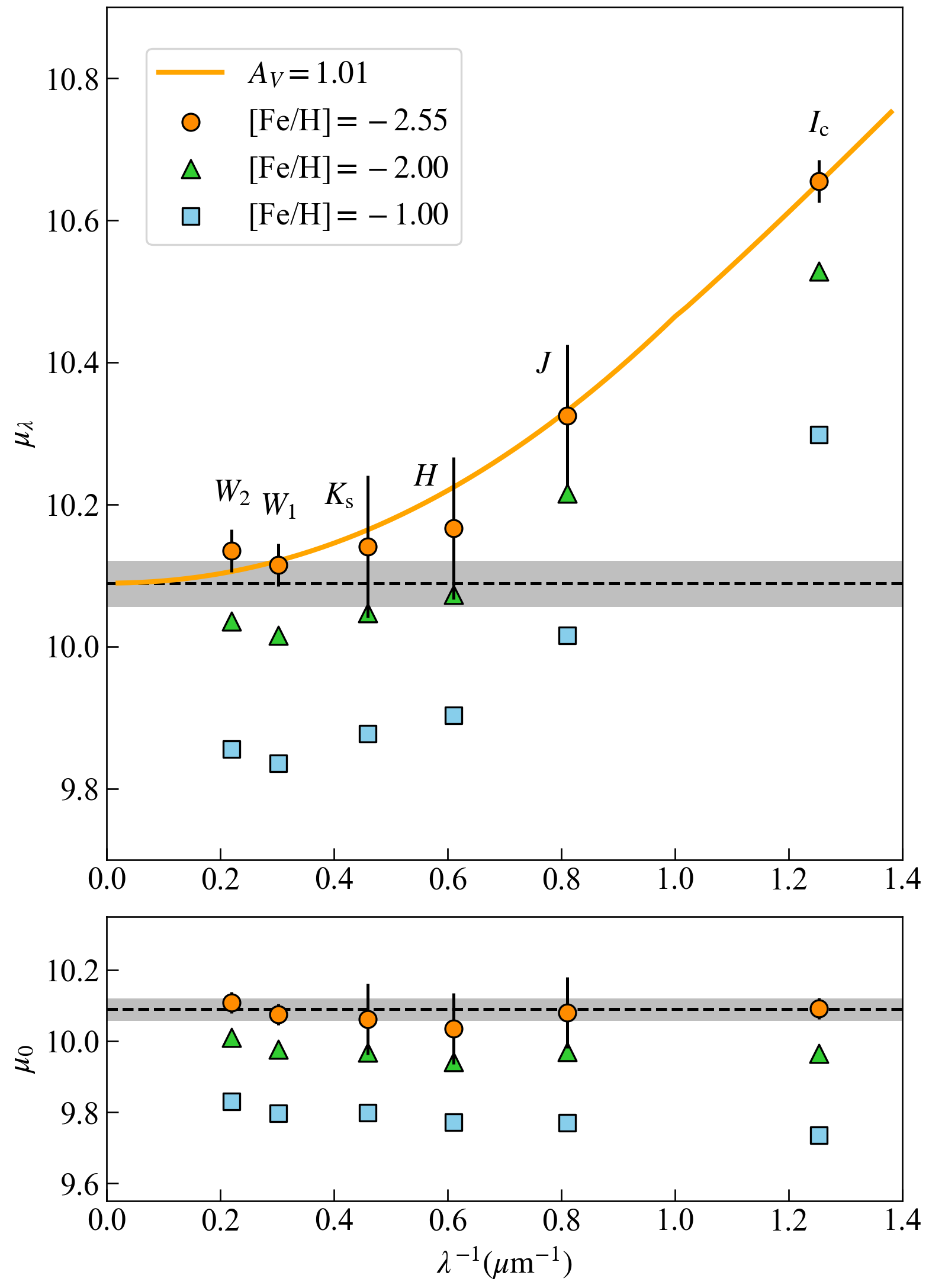}
\end{center}
\caption{
The {\it apparent} distance moduli $\mu_\lambda$ (upper panel) and
the {\it true} distance moduli $\mu_0$ (lower panel) obtained with different photometric bands.
The horizontal gray line and strip indicate
the astrometry-based distance modulus and its error obtained by
\citet{BailerJones-2021}.
The interstellar extinction corresponding to $A_V=1.01$
is used to draw the orange curve in the upper panel
and to convert $\mu_\lambda$ to $\mu_0$ at each band.
\label{fig:distance}}
\end{figure}

Combining an observed magnitude ($m_\lambda$)
and the PLZ relation (Equation~\ref{eq:PLZ}),
we can calculate the distance modulus as a function of $\FeH$,
\begin{equation}
\mu_\lambda = \mu_0 + A_\lambda = m_\lambda - M_\lambda ,
\end{equation}
where $\mu_\lambda$ and $\mu_0$ are called {\it apparent}
and {\it true} distance moduli, respectively, and
$A_\lambda$ indicates the interstellar extinction at each wavelength. 
In the upper panel of Figure~\ref{fig:distance}, the {\it apparent} 
distance moduli with different [Fe/H] are compared with each other and
also with the distance modulus corresponding to the astrometry-based distance
in \citet{BailerJones-2021}.
We adopt the extinction law obtained by \citet{Wang-2019},
i.e., $A_{\Ic} / A_V = 0.559$, $A_J/A_V = 0.243$, $A_H/A_V=0.131$,
$A_{\Ks}/A_V = 0.078$, $A_{W_1}/A_V = 0.039$, and $A_{W_2}/A_V=0.026$.

We can predict a model of $\mu_\lambda$
for a given set of $A_V$ and $\FeH$, like the one indicated
by the orange curve in Figure~\ref{fig:distance},
and we searched for the best set with the least-square method.
We used the error of 0.10\,mag for the 2MASS $JH\Ks$ and
0.03\,mag for the other bands considering
that the 2MASS data are single-epoch magnitudes.
We then obtained $A_V=1.01$
and $\FeH=-2.55$
by searching for the best set of these parameters that
makes the six-band photometry in Figure~\ref{fig:distance}
consistent with the {\it true} distance modulus
based on the astrometry-based distance.
This result indicates that our target RRL is very metal poor
($\FeH < -2$, according to the terminology in \citealt{Beers-2005}),
which is consistent with the spectroscopic analysis we present
in Section~\ref{sec:spectroscopic}.
The higher $\FeH$ would require the lower
distance as illustrated in Figure~\ref{fig:distance}. 
This estimate is subject to the systematic uncertainty and 
the intrinsic scatter of the PLZ relations given by \citet{Neeley-2019}
in addition to the uncertainty in the {\Gaia}-based distance
by \citet{BailerJones-2021}.
It is not straightforward to estimate the error in our estimate
considering various uncertainties discussed in \citet{Neeley-2019}.
We roughly estimate that the distance modulus based on 
the PLZ has the error of 0.1\,mag, which dominates
the error $\sim$0.03\,mag from the {\Gaia}-based distance,
and the error of 0.1\,mag corresponds to the error of $\sim$0.5
in $\FeH$.

The light curve shape can be used to infer the metallicity of 
RRab-type variables\citep[see][and references therein]{Mullen-2021}
but not for RRc-type ones.
Nevertheless, the period and amplitudes
indicate that this RRc star is
metal-deficient compared to typical RRLs.
Although there is a star-to-star scatter,
Fig.~7 of \citet{Sneden-2018} clearly suggests that
RRc stars with longer period tend to have lower metallicity.
Among the sample they considered, 
the relatively long period, 0.371197 days,
of the target RRL was not found among
metal-rich RRc stars ($\FeH > -1$). 
Furthermore, \citet{Fabrizio-2021} illustrated
that metal-deficient RRLs tend to have larger amplitudes
at a given period (see their Figure~7). 
Although we have no good $V$-band light curve,
the amplitudes in the {\Gaia} bands suggest
a large $V$-band amplitude, 0.55--0.65\,mag,
which is found among metal-deficient RRLs ($\FeH \lesssim -1.5$).

\section{Spectroscopic data} \label{sec:spectroscopic}
\subsection{The WINERED spectrum} \label{subsec:WINERED}
We observed the target RRL on 2015 August 15, 14:55 to 16:20 (UT) 
with WINERED attached to the 1.3\,m Araki telescope
at Koyama Observatory, Kyoto Sangyo University in Japan. 
WINERED is a near-infrared high-resolution spectrograph 
covering 0.90 to 1.35\,{$\mu$}m ($z^\prime$, $Y$, and $J$ bands)
with the resolution of $R=\lambda/\Delta\lambda = 28000$
with the WIDE mode \citep{Ikeda-2016,Ikeda-2021}.
We carried out eight 600\,s exposures, giving the total integration time
of 4800\,s, with the ABBA nodding pattern having the target within
the slit in all the exposures.
The eight exposures within {$\sim$}1.4\,hours
cover the phases 
between 0.47 and 0.63 along the pulsation cycle (Figure~\ref{fig:LCs}).

The raw spectral data were first reduced by the pipeline developed by
the WINERED team (Hamano et~al.\  in preparation). The pipeline outputs
one-dimensional spectra for individual exposures and combined spectra
along with supplementary information.
Avoiding the spectral parts with too much telluric absorption,
we consider the following echelle orders in the subsequent analysis:
43th to 48th (11560--13190\,{\AA} in the $J$ band),
51st to 57th (9760--11150\,{\AA} in the $Y$ band), and
61st (9120--9280\,{\AA} in the $z^\prime$ band).

Having the spectra from eight exposures allows us to
reject spurious noises before making the combined spectra 
and to estimate the signal-to-noise ratio (S/N) as follows.
We compared the eight one-dimensional spectra after 
the continuum normalization produced by the pipeline 
and calculated the mean ($\mu$) and the standard deviation ($\sigma$)
of the fluxes at each wavelength. We then rejected the signals
from individual exposures ($f_i$) if $|f_i-\mu|>2\sigma$
and calculated the mean and its standard error using
the accepted signals. Thus, we obtained the combined spectrum,
which is less affected by outliers, and the realistic estimates
of errors in individual pixels. 
We estimated the S/N of each echelle order by considering 
the median of the pixel-by-pixel errors in the normalized flux.
The S/N in the final spectrum per pixel ranges
from 45 to 70 except the 61st order with S/N$\sim 30$
in which the telluric absorption is rather severe.

We then performed the telluric correction 
using the synthetic telluric absorption
calculated with {\it Telfit} tool \citep{Gullikson-2014}.
We actually observed an A0V star, 29~Vul, as a telluric standard star
with the total integration time of 600\,s (two 300\,s exposures).
However, hydrogen lines are the main features we study,
and using the spectrum of a telluric standard star which shows
its own hydrogen lines would disturb 
the profiles of the hydrogen lines of our target. 
After the telluric correction, we made the continuum normalization again.

The reduced spectrum is featureless except strong hydrogen lines
and spurious noises that are mainly caused by residuals
of telluric lines and OH airglow lines \citep{Oliva-2015}.
We use the hydrogen lines to
measure the radial velocity (Section~\ref{subsec:Hlines}),
and we estimate
the upper limits of chemical abundance with the help of
theoretical synthetic spectra (Section~\ref{subsec:metallic}).

\subsection{Hydrogen lines} \label{subsec:Hlines}
\begin{figure}
\begin{center}
\includegraphics[width=0.8\hsize]{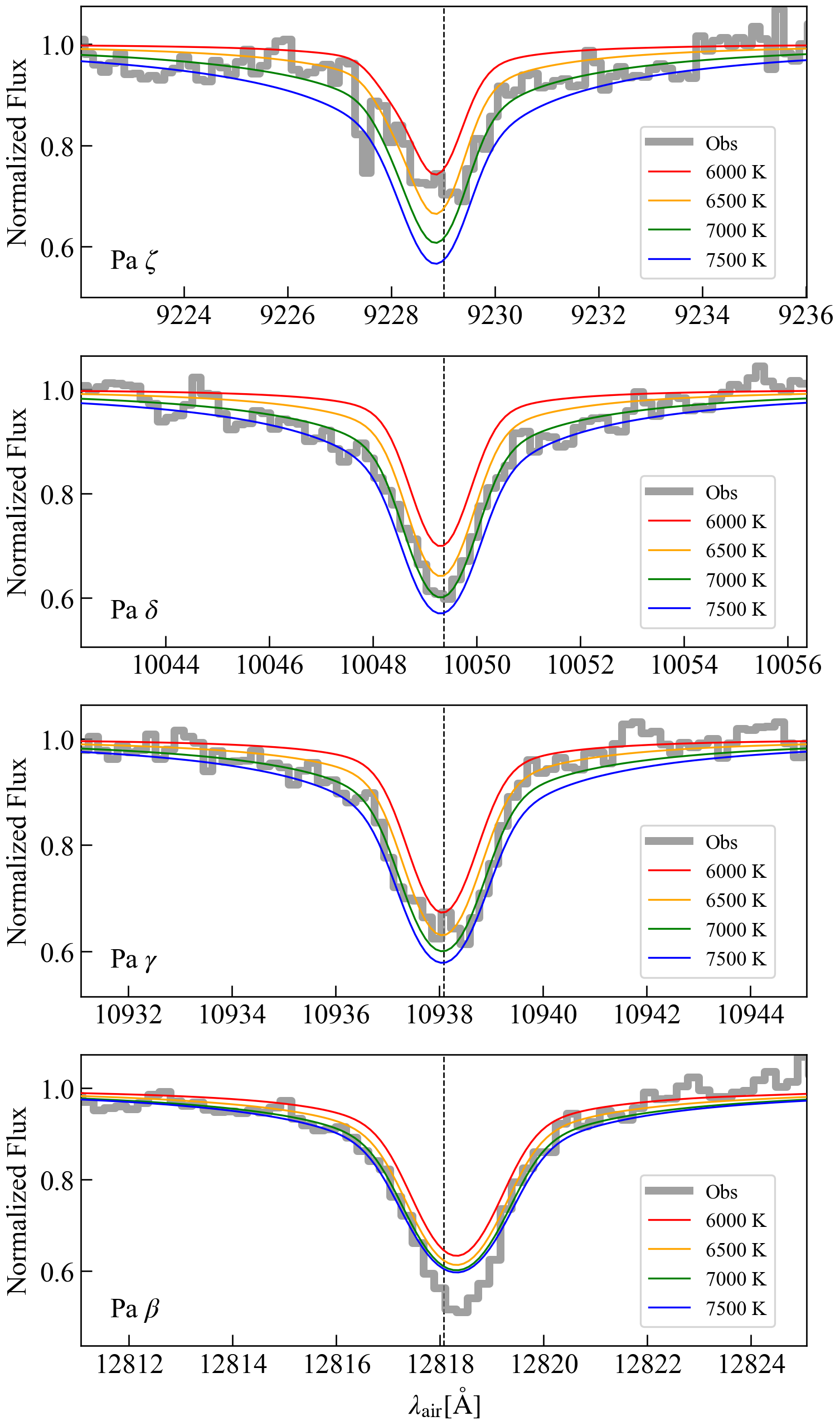}
\end{center}
\caption{
Four hydrogen lines, from Pa~$\zeta$ (top) to
Pa~$\beta$ (bottom), located within the WINERED spectrum. 
Vertical line corresponds to the air wavelength, $\lambda_{\rm obs}$,
of each line at rest.
The wavelength scale of the observed spectrum (Obs) is after
the mean radial velocity, $-85.5\,\kms$, subtracted.
Four synthetic spectra with $\log g=2.6$
and $\FeH = -2$
but with different $\Teff$ are adjusted to
the observed spectrum considering the differences between
the mean velocity and $v_i$ in Table~\ref{tab:Hlines}.
\label{fig:Hlines}}
\end{figure}

There are four hydrogen lines situated within the wavelength range  
of our interest, and we detected all of them (Figure~\ref{fig:Hlines}).
Pa~$\epsilon$ at 9545.973 is located in
the 59th order of our spectrum, but it is too much
contaminated by the telluric absorption 
in between the $z^\prime$ and $J$ bands.
We measured the central wavelength ($\lambda_{\rm obs}$)
and FWHM by fitting a Gaussian function to
20 pixels, corresponding to $\pm 50\,\kms$,
around each of the four hydrogen line.
Table~\ref{tab:Hlines} lists $\lambda_{\rm obs}$ and 
the air wavelength at rest, $\lambda_{\rm air}$, together with
the radial velocity and the FWHM. 
The velocity error in the fitting of a Gaussian is
smaller than 1\,$\kms$.
\begin{deluxetable}{crrrr}
%% \tablenum{3}
\tablecaption{Four hydrogen lines detected (Paschen series)\label{tab:Hlines}}
\tablewidth{0pt}
\tablehead{
\colhead{Line} & \colhead{$\lambda_\mathrm{air}$} & \colhead{$\lambda_\mathrm{obs}$} & \colhead{$v_i$} & \colhead{FWHM} \\
\colhead{} & \colhead{(\AA)} & \colhead{(\AA)} & \colhead{$(\kms)$} & \colhead{$(\kms)$}
}
\startdata
Pa $\zeta$  &  9229.017 &  9226.247 & $-90.0$ & $78.6$ \\
Pa $\delta$ & 10049.373 & 10046.462 & $-86.9$ & $60.8$ \\
Pa $\gamma$ & 10938.093 & 10934.966 & $-85.8$ & $62.1$ \\
Pa $\beta$  & 12818.077 & 12814.689 & $-79.3$ & $51.2$ \\
$Mean$ & & & $\mathit{-85.5}$ & \\
\enddata
\end{deluxetable}

In order to calculate the barycentric motion of the object,
however, we need to consider the pulsational effect
in addition to the heliocentric correction
(i.e., the correction taking into account the motion of the observing facility
around the Sun). 
The amplitude of radial velocity ($\Delta _\mathrm{RV}$)
is, at least approximately,
proportional to the $V$-band amplitude ($\Delta_{V}$) \citep{Sneden-2017,Magurno-2019}.
Recently, \citet{Braga-2021} thoroughly investigated 
the amplitudes and shapes of velocity curves obtained 
with different absorption lines in the optical range
and provided templates of velocity curve. 
The ratio $\Delta _\mathrm{RV}/\Delta _V$
depends on the line being measured.
In particular, H$_\alpha$ gives a large ratio, $\Delta _\mathrm{RV}/\Delta _V \simeq 107$, compared to
other Balmer lines and metallic lines, which give the ratios
${\sim}55$ with some scatter for RRc stars. 
Unfortunately, the ratios for the Paschen lines
have not been studied well. 
Therefore, we make a simple correction by 
ignoring the difference between the velocity curves
of the four Paschen lines and
assuming $\Delta _\mathrm{RV}/\Delta _V=55$.
This ratio is also consistent with the ratios reported 
by \citet{Sneden-2017} and \citet{Magurno-2019}.
The amplitudes in the {\Gaia} bands
indicate that the $V$-band semi-amplitude {$\sim$} 0.3\,mag, 
leading to the semi-amplitude of velocity ${\sim}16.5\,\kms$. 
The radial velocity of an RRc star gets most redshifted
at around the minimum phase with respect to the mean velocity
\citep[e.g.,][]{Benko-2021}. Thus, we apply the correction
of $-16.5\,\kms$ in addition to the heliocentric correction ($-2.4\,\kms$), 
resulting in the heliocentric velocity $V_\mathrm{helio}=-104.4\,\kms$ and
the velocity with respect to the local standard of rest (LSR)
$V_\mathrm{LSR}=-86.7\,\kms$.
Because of the lack of the velocity template and
because only a single-epoch velocity measurement is available,
the correction of the pulsational effect introduces
the dominant error in our estimate of the barycentric radial velocity
(or so-called gamma velocity, $V_\gamma$).
Considering the scatter of $\Delta _\mathrm{RV}/\Delta _V$
observed in different RRc stars and the line-to-line 
difference presented in \citet[][see their Fig~15]{Braga-2021},
we conservatively estimate its error to be $10\,\kms$.

\subsection{Metallic lines} \label{subsec:metallic}
We detected no metallic lines, and we cannot make
any solid estimate of the chemical abundance of the target RRL.
Instead, we estimate the upper limits of the equivalent widths (EWs)
and the corresponding limits of abundance for
the strongest absorption lines expected in the WINERED range.

\subsubsection{Spectral synthesis and stellar parameters} \label{subsubsec:synthesis}
In the subsequent analysis, we used the MOOG tool
for spectral synthesis \citep{Sneden-2012}
together with the ATLAS9 atmosphere models
extended by \citet{Meszaros-2012}. 
They provided the models with different [$\alpha$/Fe]
and [C/Fe] for a wide range of metallicity, and
we assumed [$\alpha$/Fe] and [C/Fe] are both enhanced
by $+0.3$\,dex.
Together with an atmospheric model, the spectral synthesis requires
a list of absorption lines. We considered 
Vienna Atomic Line Database \citep[VALD;][]{Ryabchikova-2015}
and the list of Mel\'endez \& Barbuy
(\citeyear{Melendez-1999}; hereinafter referred to as MB99),
and synthesized two spectra for a given atmospheric model
by using the two line lists separately.

We need stellar parameters such as 
the effective temperature
to decide which atmosphere models we use. 
The WINERED observation was carried out over {$\sim$}1.4 hours, 
within which 10-min exposures were repeated eight times).
This corresponds to a significant fraction of the pulsation cycle,
$0.473 < \phi < 0.630$ (Figure~\ref{fig:LCs}).
Nevertheless, the variation of stellar parameters
is expected to be small because the exposures were made at
around the minimum phase \citep[e.g.,][]{Govea-2014}. 
Therefore, we decided to ignore the variation of 
stellar parameters during the eight exposures. 

It is, however, not easy to obtain a precise estimate of 
the stellar parameters of our target.
In Section~\ref{subsec:PLR}, we used the PLZ relations
to estimate the interstellar reddening.
This means that we assumed that the intrinsic color of
our target is consistent with the prediction of
the PLZ relations for which only mean magnitudes are used.
On the other hand, the lack of metallic lines
prevents us from estimating the stellar parameters
with various methods
used in common spectral analyses.
Therefore, we consider the $\Teff$ and $\log g$ 
expected for RRc-type variables allowing relatively large errors.

\citet{Govea-2014} found that $\Teff$ of RRc
are concentrated in between 7000--7500\,K. 
This is consistent with $T_\mathrm{eff}$
of a larger sample of RRc stars 
in \citet{Crestani-2021b} considering the uncertainties.
At around the minimum phase, the effective temperature is expected
be at the low extreme.
We consider three temperatures, i.e., 6750, 7000, and 7250\,K, 
where we need to evaluate the effect of $\Teff$ 
on the upper limits of abundance.
Concerning other stellar parameters necessary for 
the spectral synthesis, we use the surface gravity 
$\log g=2.6$, the microturbulence $v_\mathrm{mic}=2.5\,\kms$, 
and the additional Gaussian
broadening $v_\mathrm{broad}=25.0\,\kms$
including the macroturbulence and instrumental factors.
Figure~\ref{fig:Hlines} compares the observed spectrum
with synthetic spectra with $\log g=2.6$ and $\FeH=-2$
but with four different $\Teff$ between 6000 and 7500\,K.
The relative strengths of the four Paschen lines support
the adopted temperature range. 
The constraint is, however, not very strong 
because it is hard to 
reproduce the broad profile of hydrogen lines accurately
with high-resolution echelle spectra
like the WINERED one.

\subsubsection{Upper limits of equivalent width} \label{subsubsec:EWs}
We first listed up the supposedly strongest absorption features
in the synthetic spectrum with $\Teff=7000$\,K,
$\log g=2.6$, and $\FeH=-1$ created with  
the line list of VALD or MB99. Then, we identified 
the absorption lines that form the selected features
(Tables~\ref{tab:metals-VALD} and \ref{tab:metals-MB99}).
Multiple lines can contribute to each feature.
Among the selected features, 
the one at 11659.5\,{\AA} is formed by two \ion{C}{1} lines,
and the one at 12083.5\,{\AA} by two \ion{Mg}{1} lines
according to VALD.
In addition, \ion{Mg}{2}~10914.244 and \ion{Sr}{2}~10914.887
could have formed the feature at {$\sim$}10914.5\,{\AA} together.
We excluded this mixed feature from the subsequent analysis
to avoid the blending effect on the upper limits
of abundance.
%% \tablenum{4}
\small
\begin{deluxetable}{lcrrrrc}
\tablecaption{
Upper limits of equivalent width ($W$) and [X/H] for the metallic lines selected from VALD
\label{tab:metals-VALD}}
\tablehead{
\colhead{ID} & \colhead{Species} & \colhead{$\lambda_\mathrm{air}$} & \colhead{EP} & \colhead{$\log gf$} & \colhead{$W_\mathrm{up}$} & \colhead{[X/H]$_\mathrm{up}$} \\
\colhead{} & \colhead{} & \colhead{(\AA)} & \colhead{(eV)} & \colhead{} & \colhead{(m{\AA})} & \colhead{}
}
\startdata
V01 & \ion{S}{1} & 9212.8630 & 6.524 & $0.470$ &  52 & $-1.72$ \\
V02 & \ion{Mg}{2} & 9218.2500 & 8.655 & $0.270$ &  85 & $-1.08$ \\
V03 & \ion{Si}{1} & 10585.141 & 4.954 & $0.012$ &  28 & $-1.99$ \\
V04 & \ion{C}{1} & 10683.080 & 7.483 & $0.079$ &  86 & $-2.10$ \\
V05 & \ion{C}{1} & 10685.340 & 7.480 & $-0.272$ & 103 & $-1.57$ \\
V06 & \ion{C}{1} & 10691.245 & 7.488 & $0.344$ &  72 & $-2.51$ \\
V07 & \ion{C}{1} & 10707.320 & 7.483 & $-0.411$ &  48 & $-2.05$ \\
V08 & \ion{C}{1} & 10729.529 & 7.488 & $-0.420$ &  38 & $-2.18$ \\
V09 & \ion{Si}{1} & 10827.088 & 4.954 & $0.302$ &  44 & $-2.03$ \\
V10 & \ion{Si}{1} & 10868.789 & 6.191 & $0.206$ &  39 & $-2.11$ \\
 & \ion{Si}{1} & 10869.536 & 5.082 & $0.371$ & & \\
V11 & \ion{Mg}{2} & 10914.244 & 8.864 & $0.020$ &  39 & ---\tablenotemark{a} \\
 & \ion{Sr}{2} & 10914.887 & 1.805 & $-0.638$ & & \\
V12 & \ion{C}{1} & 11658.820 & 8.771 & $-0.278$ & 144 & $-1.03$ \\
 & \ion{C}{1} & 11659.680 & 8.647 & $0.028$ & & \\
V13 & \ion{C}{1} & 11748.220 & 8.640 & $0.375$ &  61 & $-1.78$ \\
V14 & \ion{C}{1} & 11753.320 & 8.647 & $0.691$ &  46 & $-2.35$ \\
V15 & \ion{C}{1} & 11754.760 & 8.643 & $0.542$ &  42 & $-2.32$ \\
V16 & \ion{Mg}{1} & 11828.171 & 4.346 & $-0.333$ &  57 & $-1.89$ \\
V17 & \ion{Ca}{2} & 11838.997 & 6.468 & $0.312$ &  90 & $-1.66$ \\
V18 & \ion{Si}{1} & 11984.198 & 4.930 & $0.239$ &  70 & $-1.69$ \\
V19 & \ion{Si}{1} & 12031.504 & 4.954 & $0.477$ & 351 & ---\tablenotemark{b} \\
V20 & \ion{Mg}{1} & 12083.278 & 5.753 & $0.450$ &  52 & $-2.03$ \\
 & \ion{Mg}{1} & 12083.649 & 5.753 & $0.410$ & & \\
\enddata
\tablenotetext{a}{No constraint was obtained for this feature with multiple elements'  contribution.}
\tablenotetext{b}{No constraint stronger than [X/H]$\leq -1$ was given.}
\end{deluxetable}
\begin{deluxetable}{lcrrrrc}
%% \tablenum{5}
\small
\tablecaption{
Same as Table~\ref{tab:metals-VALD} but for MB99
\label{tab:metals-MB99}}
\tablehead{
\colhead{ID} & \colhead{Species} & \colhead{$\lambda_\mathrm{air}$} & \colhead{EP} & \colhead{$\log gf$} & \colhead{$W_\mathrm{up}$} & \colhead{[X/H]$_\mathrm{up}$} \\
\colhead{} & \colhead{} & \colhead{(\AA)} & \colhead{(eV)} & \colhead{} & \colhead{(m{\AA})} & \colhead{} 
}
\startdata
M01 & \ion{C}{1} & 10123.87 & 8.54 & $-0.09$ &  51 & $-1.44$ \\
M02 & \ion{Si}{1} & 10585.14 & 4.95 & $-0.06$ &  28 & $-1.92$ \\
M03 & \ion{C}{1} & 10683.09 & 7.48 & $0.03$ &  86 & $-2.05$ \\
M04 & \ion{C}{1} & 10685.36 & 7.48 & $-0.30$ & 103 & $-1.54$ \\
M05 & \ion{C}{1} & 10691.26 & 7.49 & $0.28$ &  72 & $-2.45$ \\
M06 & \ion{C}{1} & 10707.34 & 7.48 & $-0.41$ &  48 & $-2.05$ \\
M07 & \ion{C}{1} & 10729.54 & 7.49 & $-0.46$ &  38 & $-2.14$ \\
M08 & \ion{Si}{1} & 10749.39 & 4.93 & $-0.21$ &  46 & $-1.49$ \\
M09 & \ion{Si}{1} & 10827.10 & 4.95 & $0.23$ &  44 & $-1.95$ \\
M10 & \ion{Si}{1} & 10869.54 & 5.08 & $0.36$ &  39 & $-2.09$ \\
M11 & \ion{Mg}{2} & 10914.24 & 8.86 & $0.00$ &  39 & ---\tablenotemark{a} \\
 & \ion{Sr}{2} & 10914.88 & 1.80 & $-0.59$ & & \\
M12 & \ion{C}{1} & 11658.85 & 8.77 & $-0.36$ & 144 & ---\tablenotemark{b} \\
 & \ion{C}{1} & 11659.70 & 8.65 & $-0.07$ & & \\
M13 & \ion{C}{1} & 11748.24 & 8.64 & $0.40$ &  61 & $-1.81$ \\
M14 & \ion{C}{1} & 11753.32 & 8.65 & $0.69$ &  46 & $-2.33$ \\
M15 & \ion{C}{1} & 11754.79 & 8.64 & $0.51$ &  42 & $-2.29$ \\
M16 & \ion{Mg}{1} & 11828.19 & 4.35 & $-0.50$ &  57 & $-1.72$ \\
M17 & \ion{Ca}{2} & 11838.99 & 6.47 & $0.24$ &  90 & $-1.59$ \\
M18 & \ion{Ca}{2} & 11949.76 & 6.47 & $-0.04$ &  43 & $-1.92$ \\
M19 & \ion{Si}{1} & 11984.23 & 4.93 & $0.12$ &  70 & $-1.57$ \\
M20 & \ion{Si}{1} & 12031.53 & 4.95 & $0.24$ & 351 & ---\tablenotemark{b} \\
\enddata
\tablenotetext{a}{No constraint was obtained for this feature with multiple elements'  contribution.}
\tablenotetext{b}{No constraint stronger than [X/H]$\leq -1$ was given.}
\end{deluxetable}

For each selected feature, 
we evaluated the upper limits of the EW as follows.
First, we calculated the weighted mean and its error of 
the pixel counts within $\pm 150\,\kms$ around each line
but with the pixels within $\pm 25\,\kms$ of
the features in Tables~\ref{tab:metals-VALD} and \ref{tab:metals-MB99} excluded from the calculation.
The weighted mean is considered as the local continuum level, $f_\mathrm{c}$ (with the error, $e_\mathrm{c}$),
in the wavelength range around the feature.
If the continuum normalization for each order were
perfect in the spectral reduction (Section~\ref{subsec:WINERED}),
$f_\mathrm{c}$ should be 1,
and it is actually consistent with 1 within the error
in most cases.
Then, we obtained the EW ($W$) and its error ($E_W$) by
\begin{eqnarray}
W &=& \sum_{i=1}^{n} \left(1-\frac{f_i}{f_\mathrm{c}} \right) \Delta\lambda_i \\
E_W &=& \frac{1}{f_\mathrm{c}} \sqrt{\sum_{i=1}^n \sigma_i^2 (\Delta\lambda_i)^2} + \frac{e_c}{f_\mathrm{c}^2} \sum_{i=1}^n f_i \Delta\lambda_i
\end{eqnarray}
where the sum was taken over the $n$ pixels
($1\leq i \leq n$, with the flux $f_i$ and the noise $\sigma_i$ at each pixel)
within $\pm 25\,\kms$ around the line center
and $\Delta\lambda_i$ indicates the width of each pixel
in the unit of m{\AA}.
We estimated the upper limit of EW by
\begin{eqnarray}
W_\mathrm{up} = \left\{ \begin{array}{ll}
W+3\,E_W & ({\rm if} ~W\geq 0) \\
3\,E_W & ({\rm if} ~W<0) 
\end{array}
\right.
\end{eqnarray}
for each feature (Tables~\ref{tab:metals-VALD} and \ref{tab:metals-MB99}).

\subsubsection{Upper limits of elemental abundance} \label{subsubsec:metals}
We estimated the upper limits of chemical abundance
based on the upper limits of EW.
This was done with the help of synthetic spectra.
For each feature in Tables~\ref{tab:metals-VALD} and \ref{tab:metals-MB99},
except the mixed feature of \ion{Mg}{2} and \ion {Sr}{2} at 10914.6\,{\AA},
we calculated the EWs in the synthetic spectra
with different abundances of each species over $-3 \leq \FeH \leq -1$.
This enables us to draw the curve of growth and estimate the upper limit
which corresponds to % the upper limit of the observational EW.
$W_\mathrm{up}$.
We estimated the abundance upper limits with the models at
three different temperatures (6750, 7000, and 7250\,K), and
took the highest upper limit of the three as the final estimate,
[X/H]$_\mathrm{up}$ in Tables~\ref{tab:metals-VALD} and \ref{tab:metals-MB99},
based on the given feature.
The maximum metallicity of the synthetic spectra we considered is $-1$\,dex.
If $W_\mathrm{up}$ is larger than
the maximum EW that we found with the synthetic spectra,
we give no constraint on the abundance for a given line.
\begin{figure}
\begin{center}
\includegraphics[width=0.8\hsize]{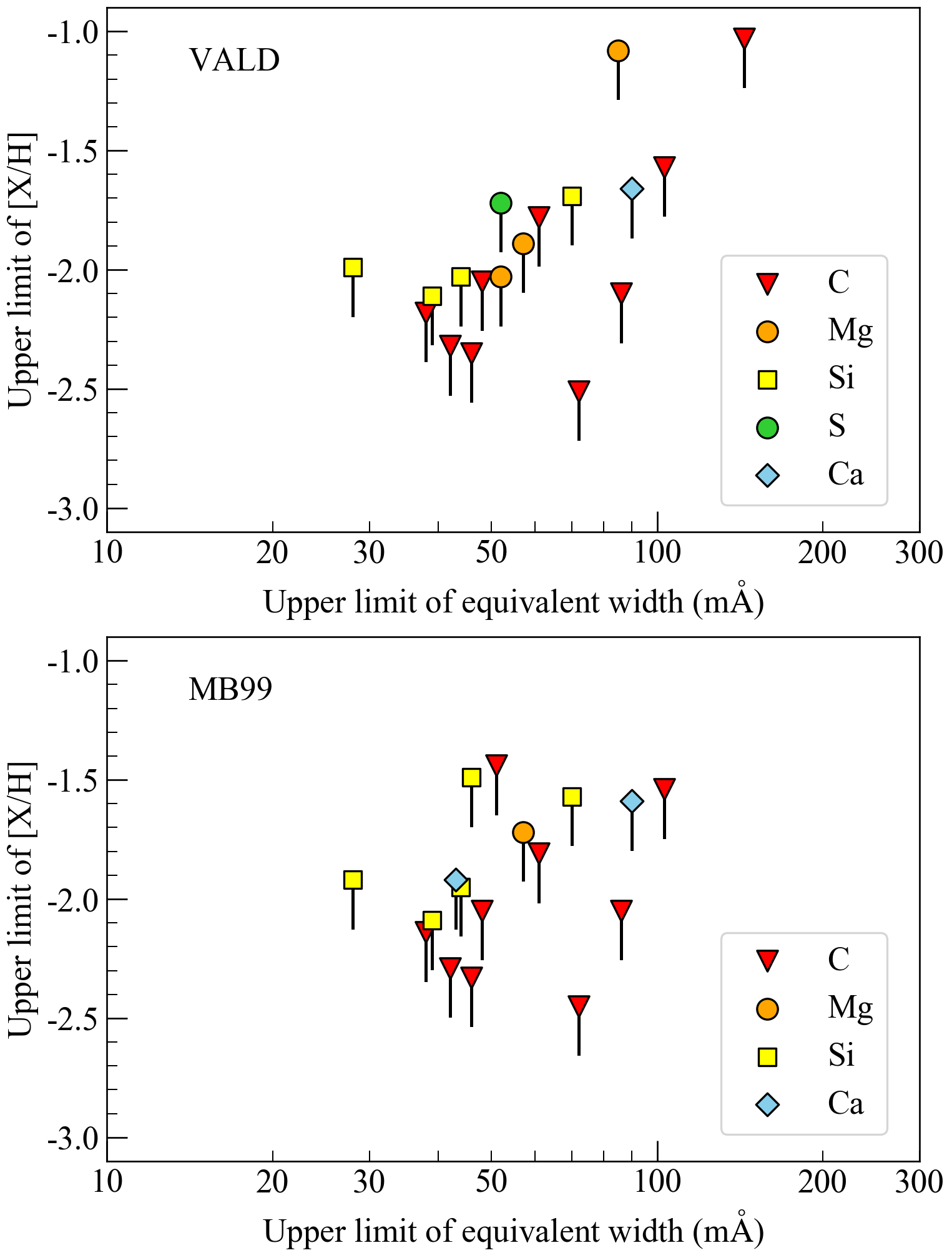}
\end{center}
\caption{
Upper limits of [X/H] given by individual lines in Table~\ref{tab:metals-VALD} (VALD) and those in Table~\ref{tab:metals-MB99} (MB99).
\label{fig:upper_limits}}
\end{figure}

Figure~\ref{fig:upper_limits} plots the upper limits of abundance.
For most of the lines included in both
VALD (Table~\ref{tab:metals-VALD}) and 
MB99 (Table~\ref{tab:metals-MB99}), the $\log gf$ in the two lists
agree with each other within 0.1\,dex, and the difference
in the line list is not important for the upper limits
in Figure~\ref{fig:upper_limits}. 
Neutral carbon lines give the strongest constraints
in terms of [X/H], i.e., [C/H]$\lesssim -2.5$.
A couple of lines of other elements (Mg, Si, and Ca) 
indicate [X/H]~$\lesssim -2$. 
We have no direct constraint on [Fe/H] because
no iron line in the WINERED wavelength range
is expected to be so strong as the lines in  
Tables~\ref{tab:metals-VALD} and \ref{tab:metals-MB99}.
There have been several reports of 
carbon-enhanced stars among metal-deficient RRLs
\citep[e.g.,][]{Preston-2006,Kinman-2012,Kennedy-2014}.
In contrast, \citet{Andrievsky-2020} found [C/Fe]$< 0$
for a few RRLs with $-1.7\leq \FeH \leq -1.2$
based on the non-local-thermodynamic-equilibrium (NLTE) analysis.
We simply give the upper limit of $-2.5$\,dex.
and use it in the subsequent discussion. 
According to the terminology defined by \citet{Beers-2005},
the target RRL is a very metal-poor star ($\FeH<-2$) 
if it is not an extremely metal-poor star ($\FeH<-3$).

\section{Discussion} \label{sec:discussion}
Combining the radial velocity estimated with
the four Paschen lines (Section~\ref{subsec:Hlines})
with the {\Gaia}'s distance and proper motions,
the six-dimensional information (i.e., the position
and space velocity) is available for our target. 
We computed the target's orbital and kinematic properties
by taking into account the observational uncertainties.
We used the AGAMA package \citep{Vasiliev-2019}
with the Galactic constants adopted from \citet{Zinn-2020}:
the distance to the Galactic center $R_0=8.2$\,kpc,
the velocity of the local standard of rest (LSR)
$v_\mathrm{LSR}=232\,\kms$, and
the solar velocity with respect to the LSR
being $(U_\odot, V_\odot, W_\odot)=(-11.1, 12.24, 7.25)\,\kms$ \citep{Schonrich-2010}.
We used the Galactic potential 
called MWPotential2014 available in
the galpy library \citep{Bovy-2015}, which is composed
of three axisymmetric potentials
for the spherical power-law bulge with an exponential cut-off,
a Navarro-Frenk-White halo potential, and
a Miyamoto-Nagai disk. 

We randomly drew 10000 samples from the error distribution
of the position and velocity and integrated the orbit forward
in time for a long enough period of time (100 Gyr).
The average and standard deviation of the orbital parameters from individual Monte-Carlo samples 
are given in Table\,\ref{tab:orbit}. 
The eccentricity is defined by $e=(\rmax-\rmin)/(\rmax+\rmin)$.
The positive azimuthal velocity ($v_\theta$) and
angular momentum ($L_Z$) correspond
to the prograde rotation.
With the maximum height $z_\mathrm{max}=1.18$\,kpc
from the Galactic plane, the orbit of the target RRL
is accommodated within the stretch of the thick disk
whose vertical scale length is about 0.9\,kpc
\citep{BlandHawthorn-2016}.
%% \tablenum{6}
\begin{deluxetable}{lc}
\scriptsize
\tablecaption{Kinematic properties of the target RR~Lyr star\label{tab:orbit}}
\tablehead{
\colhead{Parameter} & \colhead{Value}
}
\startdata
\multicolumn{2}{c}{\it Input parameters} \\
$D_0$---geometric distance & $1.042 \pm 0.015$ ~(kpc) \\
$v_\mathrm{helio}$---radial velocity & $-104 \pm 10$ ~($\kms$) \\
$\mu_\alpha \cos \delta$---proper motion along $\alpha$ & $11.76 \pm 0.01$ ~(mas~yr$^{-1}$) \\
$\mu_\delta$---proper motion along $\delta$ & $-5.09 \pm 0.02$ ~(mas~yr$^{-1}$) \\
\multicolumn{2}{c}{\it Output parameters} \\
$\rmin$---pericenter distance & $3.71 \pm 0.38$ ~(kpc) \\
$\rmax$---apocenter distance & $8.36 \pm 0.01$ ~(kpc) \\
$\zmax$---maximum height & $1.18 \pm 0.01$ ~(kpc) \\
$e$---eccentricity & $0.39 \pm 0.04$ \\
$v_\mathrm{R}$---radial velocity & $52.9 \pm 2.2$ ~($\kms$) \\
$v_\theta$---azimuthal velocity & $143.5 \pm 9.8$ ~($\kms$) \\
$v_Z$---vertical velocity & $-53.6 \pm 0.3$ ~($\kms$) \\
$v_\mathrm{pec}$---peculiar velocity & $118.7 \pm 8.4$ ~($\kms$) \\
$E_\mathrm{tot}$---total orbital energy & $-118.9 \pm 1.3$ ~(10$^3$~km$^2$~s$^{-2}$) \\
$L_Z$---azimuthal angular momentum & $1135 \pm 77$ ~(kpc~km~s$^{-1}$) \\
\enddata
\tablecomments{The given errors are the standard deviation observed in our Monte-Carlo calculation and do not include the systematic errors. The velocity and proper motion of the {\it input parameters} are given with respect to the Sun, while the velocities of the {\it output parameters} are given with respect to the Galactic center.}
\end{deluxetable}

\begin{figure}
\begin{center}
\includegraphics[width=0.8\hsize]{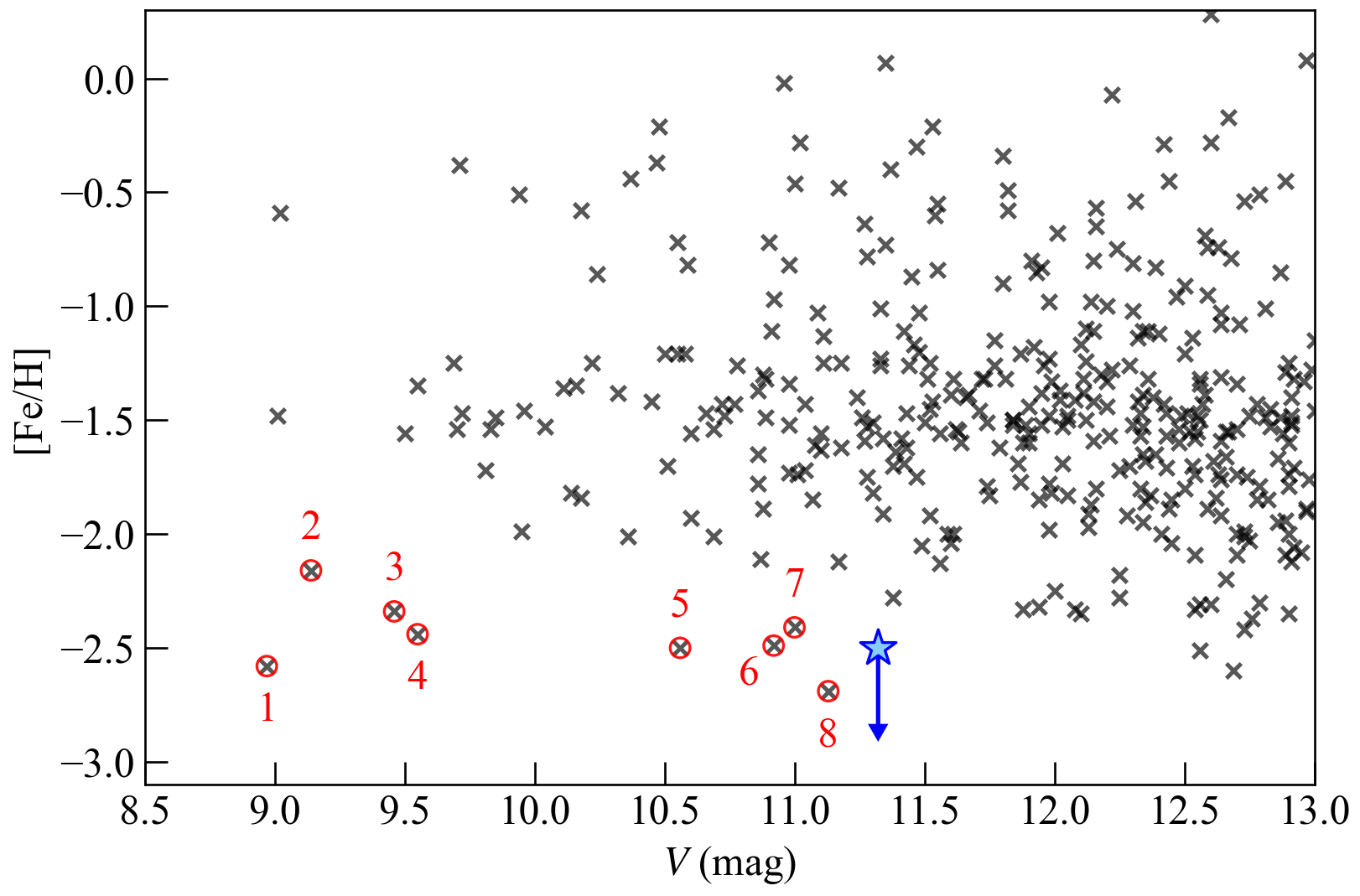}
\end{center}
\caption{
The metallicities and $V$-band magnitudes
of the target RRL (blue) and the known RRLs compiled
by \citet{Zinn-2020}.
The metallicity upper limit of
the target RRL is indicated by the star symbol accompanied by arrow. 
Red circles highlight the bright and metal-deficient objects
we selected for comparisons (see text):
1$=$MT~Tel, 2$=$V338~Pup, 3$=$RZ~Cep, 4$=$X~Ari, 
5$=$V701~Pup, 6$=$UY~Boo, 7$=$TV~Boo, and 8$=$ASAS~200431$-$5352.3. 
\label{fig:comp_Z20_1}}
\end{figure}

\begin{figure*}
\begin{center}
\includegraphics[width=0.8\hsize]{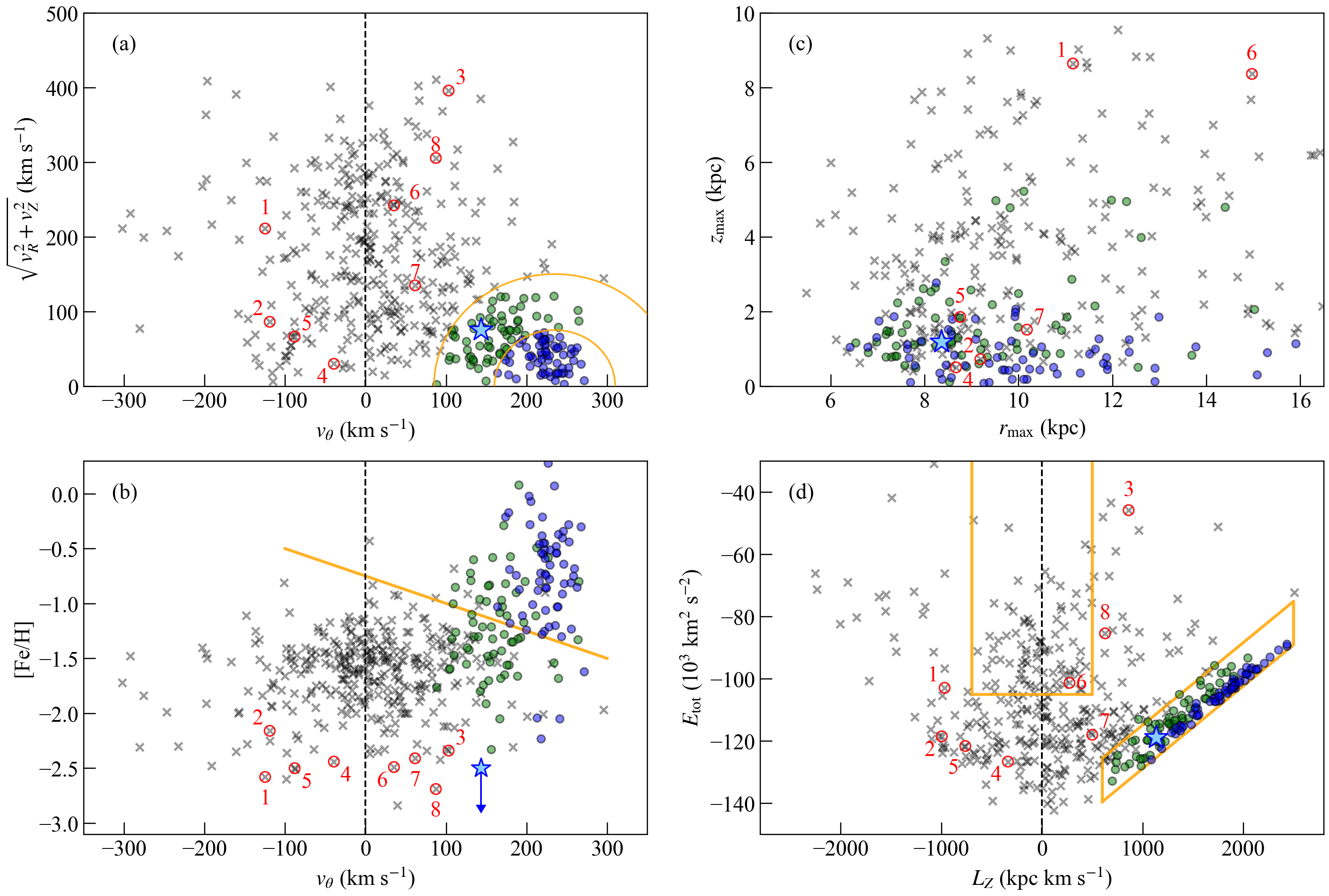}
\end{center}
\caption{
Properties of the target RRL in comparison with
the known RRLs compiled by \citet{Zinn-2020}.
The target RRL is indicated by the star symbol, and
the arrow in panel (b) means that its metallicity
is given as the upper limit.
The orange semi-circles in panel (a) indicate
the peculiar velocities $v_\mathrm{pec}$ of 
75 and 150\,$\kms$.
The RRLs with $v_\mathrm{pec}$ smaller than 75\,$\kms$, 
are indicated by blue circles, 
and those with $v_\mathrm{pec}$ between 75 and 150\,$\kms$
by green circles, and the other RRLs from \citet{Zinn-2020}
are indicated by gray crosses.
The same bright and metal-deficient RRLs as in
Figure~\ref{fig:comp_Z20_1} are indicated by red circles
with the number IDs labelled. 
The orange line in panel (b) indicates
the threshold, $v_\theta = -400\,\FeH - 300$,
used for selecting disk RRLs by \citet{Layden-1996},
while the regions enclosed by orange lines
in panel (d) indicate
the ``Plume'' (center) and the ``disk'' (right). 
\label{fig:comp_Z20_2}}
\end{figure*}

In Figures~\ref{fig:comp_Z20_1} and \ref{fig:comp_Z20_2}, we compare the properties of
our target RRL with 463 RRLs compiled by \citet{Zinn-2020},
but we re-calculated their parameters except [Fe/H] and
$V$-band magnitudes.
We combined the radial velocities adopted from \citet{Zinn-2020}
with the astrometric data from the {\Gaia} EDR3
and the EDR3-based distances from \citet{BailerJones-2021}
to calculate the current positions ($X$, $Y$, and $Z$),
the velocities ($v_R$, $v_\theta$, and $v_Z$),
the apocenter distance ($\rmax$),
the maximum height ($\zmax$),
the angular momenta ($L_Z$), and the total orbital energies 
($E_\mathrm{tot}$, the sum of kinetic and potential energies).
The total orbital energies show a systematic offset,
$\sim 7 \times 10^4$\,km$^2$\,s$^{-2}$, between our calculation and
that of \citet{Zinn-2020}
because of the difference in the Galactic potential.
The distributions in the other parameters do not show
such systematic offsets but the parameters of individual objects
are expected to be improved by using the {\Gaia} EDR3.

Figure~\ref{fig:comp_Z20_1} includes 360 RRLs with $V<13$, 
while our RRL is located at $V=11.32$, 
obtained in Section~\ref{subsec:other-photo}
using the {\Gaia} photometry.
We highlight eight bright and metal-deficient ($\FeH\lesssim -2.3$) objects
selected within Figure~\ref{fig:comp_Z20_1}
and list their names in the caption.
Three of them (V338~Pup, X~Ari, and UY~Boo) are fundamental-mode
pulsators (i.e., RRab type), while the other five are
first-overtone pulsators (RRc, same as the target RRL).
The metallicities of the eight RRLs have been measured and reported
recently in 
\citet{Beers-2014}, \citet{Sneden-2017}, \citet{Andrievsky-2018}, \citet{Chadid-2017}, and \citet{Zinn-2020}.
The brighter four RRLs (No.~1--4) are located at 0.4--0.6\,kpc from the Sun,
while the fainter four are further, at 0.95--1.3\,kpc,
according to \citet{BailerJones-2021}.
The distance to our target RRL is among the more distant ones,
and its metallicity is as low as the eight RRLs. 

The panel (a) of Figure~\ref{fig:comp_Z20_2} presents
the Toomre diagram, in which the semi-circle contours
indicate constant peculiar velocities,
\begin{equation}
v_\mathrm{pec} = \sqrt{v_R^2+(v_\theta-v_\mathrm{LSR})^2+v_Z^2} ,
\end{equation}
with respect to the LSR corresponding to 75
and 150\,$\kms$. The panel (b) presents that
the RRLs with relatively small $v_\mathrm{pec}$
tend to be metal-rich \citep{Layden-1995}.
In Figure~\ref{fig:comp_Z20_2}, except the panel (c),
halo stars show more-or-less symmetric distributions
of $v_\theta$ and $L_Z$, and they overlap with
the disk component at around the position of our target.
Bright and metal-deficient RRL highlighted in Figures~\ref{fig:comp_Z20_1} and \ref{fig:comp_Z20_2}
can be classified as halo objects according to their $v_\mathrm{pec}$.
The motion of our RRL star deviates significantly
from the Galactic rotation with $v_\mathrm{LSR} = 232\,\kms$, but
the star is, with the peculiar velocity of 116.6\,$\kms$,
indistinguishable from thick-disk stars in regards to the kinematics
(Figure~\ref{fig:comp_Z20_2}a).
Our RRL may still belong to the halo
and its motion is at the prograde-side tail of
halo orbits. Two metal-deficient bright RRLs highlighted,
V338~Pup and V701~Pup (Nos.~2 and 5), are located
at around the opposite point on the retrograde side. 

The panel (c) plots $z_\mathrm{max}$ against $r_\mathrm{max}$
estimated with the orbit calculation. Like our RRL,
the orbits of the four bright, metal-deficient RRLs highlighted
are within the stretch of the thick disk, although three of them show retrograde motion. 
In addition, the four RRLs have larger eccentricity
than our RRL: 0.53 (V338~Pup), 0.60 (V701~Pup), 0.79 (TV~Boo),
and 0.83 (X~Ari) in contrast to 0.39 of our RRL.
Nevertheless, the kinematics of all these
very metal-poor RRLs, including our target RRL, is 
consistent with that of similarly metal-deficient
stars investigated by \citet{Chiba-2000}, 
who concluded that the disk population is negligible
at $\FeH \leq -2.2$. 
In a recent study on a large sample of stars towards
the Galactic anticenter, \citet{FernandezAlvar-2021} detected metal-poor stars
belonging to the thin disk well down to $\FeH \simeq -2$ but the situation
of more metal-deficient stars was not conclusive.

The panel (d) of Figure~\ref{fig:comp_Z20_2} plots
$E_\mathrm{tot}$ against $L_Z$.
\citet{Zinn-2020} found that, in addition to
the RRLs that may be related to moving groups
like the Helmi stream \citep{Helmi-1999},
there are two major groups, i.e., the `disk' RRLs
with prograde rotation and the `Plume' RRLs
with $L_Z \sim 0$. These two groups of RRLs were
also found by \citet{Prudil-2020} and \citet{Iorio-2021}.
The Plume structure of
halo stars was discovered by \citet{Dinescu-2002},
and recent studies based on {\Gaia} data identified 
the very prominent feature 
called {\it Gaia Enceladus} or {\it Gaia Sausage} 
\citep{Belokurov-2018,Helmi-2018}.
This prominent feature is considered to originate
from an accreted galaxy that
contributed many halo objects, including globular clusters
such as $\omega$~Cen after the major merger with
the Milky Way \citep{Belokurov-2018,Helmi-2018}.
\citet{Zinn-2020} found that 
Plume RRLs include fewer objects with $\FeH < -2$,
and, among the eight highlighted objects,
UY~Boo (No.~6) is the only one located within
the Plume region in Figure~\ref{fig:comp_Z20_2}(d). 

Our RRL does not belong to the Plume but
is associated with, or at least closer to,
the disk populations as discussed above. 
There are accumulating reports and discussions on
the presence of 
metal-deficient stars in the thick disk
\citep{DiMatteo-2020,Sestito-2020,Limberg-2021}.
The latter two authors used large samples of 
more than 1000 candidates of metal-deficient stars
and found 
a limited but significant fraction of stars with
disk orbits ($\zmax < 3$\,kpc and $6\lesssim \rmax\lesssim 13$\,kpc) and low metallicity ($\FeH<-2.5$).
Our target RRL may belong to the same population. 
There are a few scenarios to explain the metal-deficient
disk population \citep[see, e.g.,][]{Sestito-2020}. 
Such a population could compose the ancient disk of the Galaxy
present before the severe merger that created the Plume
(or the {\it Gaia-Sausage-Enceladus}) structure around 10\,Gyr ago. 
The formation of the stars contributing to the ancient disk
may be {\it in situ} (within the pre-existent disk) or external.
Alternatively, metal-deficient stars with external origins could
be quietly merged into the Galactic thick disk even after
the severe merger \citep{Gomez-2017,Karademir-2019}.
If the membership to the thick disk is confirmed, 
the target RRL would be a unique object representing 
the population of the metal-deficient thick disk. 
Otherwise, the target may be giving a caution for contamination of
halo stars to the disk population. 

\section{Concluding remarks} \label{sec:conclusion}
We presented photometric and spectroscopic analysis on
an RRL, KISOJ\,201241.60$+$321242.4 or HD\,331986,
located in the Galactic plane
at 1\,kpc from the Sun. Although this star
was found to be an RRL by some previous surveys and 
is bright ($V\sim 11.3$), no study investigated its detailed characteristics.
We confirmed its classification as an RRc-type variable
and discovered that it is a very metal-poor star. 
The near-infrared spectrum taken
with the WINERED spectrograph covering 0.9--1.35\,{$\mu$}m
shows only hydrogen lines but no metallic lines. 
We estimated the upper limit of metallicity to be $\FeH=-2.5$.
This upper limit is consistent with the metallicity
inferred from the period--luminosity--metallicity relation,
although there remains a large uncertainty, $\sim$0.5\,dex,
in the latter estimate.
We conclude that the object is among the 
known RRLs with the lowest metallicity, around $-3.0$ to $-2.5$\,dex
\citep{Hansen-2011,Crestani-2021a}.

This RRL is located within the thick disk and its kinematics is 
consistent with that of thick-disk objects, which
makes it an even more interesting object. 
While RRLs with $\FeH \gtrsim -1$
tend to have disk-like orbits
\citep{Layden-1995,Layden-1996,Prudil-2020,Zinn-2020},
metal-deficient RRLs have been regarded as halo objects. 
Recent studies \citep[e.g.,][]{Sestito-2020,Limberg-2021} found 
the presence of metal-deficient stars (but not RRLs) in the thick disk, 
which has a large impact on our understanding of the Galactic formation. 
Finding the origin of the target RRL would give an essential insight into
the early history of the Galaxy.

Detailed elemental abundances are crucial for disclosing the origins of stars.
However, without any metallic lines detected, we have no clues to
the abundance pattern of the target. 
It is of vital importance to make
follow-up spectroscopic observations
in the optical range, in which 
much stronger lines of various elements are present
\citep[e.g.,][]{Hansen-2011,Crestani-2021b}.
The origin of our target may be revealed
by comparing its abundance pattern
with those of
RRLs and other stars with similarly low metallicity
in different groups including the halo and the thick disk.
Furthermore, the census of RRLs in 
the Galactic plane region has been limited,
and future surveys including the {\Gaia}
observations would reveal more metal-deficient
RRLs that are constrained in the disk. 
A larger sample of such objects and follow-up
observations would enable us to understand
the initial environment and formation
of the ancient Galactic disk. 

\begin{acknowledgments}
We acknowledge useful comments from the anonymous referee.
We are grateful to the staff of 
Kiso Observatory, the University of Tokyo, and 
Koyama Astronomical Observatory, Kyoto Sangyo University,
for their support during our observations. 
This study is supported by JSPS KAKENHI (grant
Nos.~23684005, 26287028, and 18H01248). The WINERED was developed
by the University of Tokyo and the Laboratory of Infrared High-resolution
spectroscopy (LiH), Kyoto Sangyo University under the
financial supports of KAKENHI (Nos.~16684001, 20340042, and
21840052) and the MEXT Supported Program for the Strategic
Research Foundation at Private Universities (Nos. S0801061 and S1411028). 
Two of us are financially supported by JSPS Research Fellowship for Young
Scientists and accompanying Grant-in-Aid for JSPS Fellows, MJ
(No. 21J11301) and DT (No. 21J11555). DT also acknowledges
financial support from Masason Foundation.
KH is supported by JSPS KAKENHI Grants (Nos.~21K13965 and 21H00053).
VB acknowledges the ﬁnancial support of the Istituto
Nazionale di Astroﬁsica (INAF), Osservatorio Astronomico di
Roma, and Agenzia Spaziale Italiana (ASI) under contract to
INAF: ASI~2014-049-R.0 dedicated to SSDC.
This research has made use of the SIMBAD database 
and the VizieR catalog access tool,
provided by CDS, Strasbourg, France (DOI:~10.26093/cds/vizier).
The original description of the VizieR service was published in 2000 (A\&AS 143, 23).
\end{acknowledgments}

% \vspace{5mm}
% \facilities{HST(STIS), Swift(XRT and UVOT), AAVSO, CTIO:1.3m,
% CTIO:1.5m,CXO}

\software{AGAMA \citep{Vasiliev-2019}, IRAF \citep{Tody-1986,Tody-1993}, MOOG \citep[2019 November version;][]{Sneden-2012}, WINERED pipeline (S. Hamano et al. 2021, in preparation).}

%% Include this line if you are using the \added, \replaced, \deleted
%% commands to see a summary list of all changes at the end of the article.
%\listofchanges

\end{document}